Whitepaper
January 2019

Centre of Excellence DNS SECURITY

# An Intrusion Using Malware and DDNS

Written by: Gopinath Palaniappan, Dr. Balaji Rajendran, Dr. S Sangeetha, Dr. Kumari Roshni V S

**Contents:**

*An intrusion executed by Hidden Cobra team using RAT Joanap and SMB Brambul malware with support of DDNS and proxy services*

## Introduction

**Domain Name System (DNS)** is a simple *domain name* to *IP address* translation service and vice-versa. The IP addresses associated with Web Servers, FTP Servers, Mail Servers, etc., are generally static in nature, meaning they are fixed such that the *name* to *IP address* relationship is constant. These *name-IP address* mapping records in the DNS tend to be manually assigned by the DNS service administrators.

In large corporate networks or home networks, when the machines boot, they are assigned dynamic IP addresses by DHCP service (Dynamic Host Configuration Protocol) usually configured in a home router. Such networks mostly run their own DNS service, with the ability to update the DNS records as a machine boots or as its IP address changes. A DNS service in which the user has the flexibility of controlling the *name-IP address* mapping is known as ***Dynamic DNS***.

This whitepaper captures the details of the technical alert (**TA18-149A**) dated 29th May 2018 issued by the US-CERT (United States Computer Emergency Readiness Team). In the alert, the US-CERT had warned of two pieces of malware and claimed as being used by the North Korea's **Hidden Cobra** hacker team to attack and access networks worldwide. The two pieces of malware were: **Remote Access Trojan (RAT) Joanap** and **Server Message Block (SMB) Brambul worm** malware, which belong to the prolific Advanced Persistent Threat (APT) group of malware [1].

Few resources even claim that the Hidden Cobra actors have likely been using both Joanap and Brambul malware since atleast 2009 to target networks and machines spread across the globe. The US government has found Joanap on 87 compromised networks in about 17 countries. The Hidden Cobra actors had planned their act of intrusion and exploitation using the malwares by setting-up an infrastructure comprising services registered with Dynamic DNS services and concealing their own network behind proxy services. Few members of the Hidden Cobra team were also involved in authoring the globally popular WannaCry2.0 ransomware in 2017.

The Brambul is a malicious Windows 32-bit Server Message Block (SMB) worm

The Destover dropper malware dropped the Brambul worm and RAT Joanap on the victims' machines

The RAT Joanap facilitates communication and botnets management as per commands from a C&C server

## The SMB Brambul Worm

***Brambul Worm*** is a malicious **Windows 32-bit Server Message Block (SMB)** worm that functions as a service *dynamic link library (dll) file* or a *portable executable file* often dropped and installed onto victims' network by a **dropper malware**. The Server Message Block (SMB) is a method that Microsoft systems use to share files on a network. Brambul Worm, when executed attempts to establish contact with victim systems and IP addresses on victims' local subnets. If successful, the malware targets insecure or unsecured user accounts and spreads through poorly secured network shares and attempts to gain unauthorized access via the SMB protocol (ports 139 and 445) by launching brute-force password attacks using a list of embedded passwords. Additionally, the malware communicates information about victim's systems (IP address, host name, username, password etc.) to Hidden Cobra actors using malicious email addresses embedded within it and also generates random IP addresses for further attacks. The Hidden Cobra actors then use the system information to remotely access a compromised system via the SMB protocol.

In brief, the Brambul malware can perform following functions for remote operations: (1) harvesting system information, (2) network propagation using SMB, (3) brute forcing SMB login credentials, (4) generating SMTP email messages, (5) accepting command-line arguments, and (6) generating and executing a suicide script [1].

A dropper belongs to the category of trojans, which embeds a malware within it and installs itself and the malware carried when it gets downloaded. Meredrop, and Destover are popular *dropper malware*. The ***Destover malware*** also contained a wiper to overwrite or erase system executables or program files, rendering infected computers inoperable; and it also could connect to a webserver whose IP address was hard-coded within it and download web pages to victim machines.

## The RAT Joanap malware

***Joanap malware*** is a **Remote Access Trojan (RAT)** that is capable of receiving multiple commands from a remote Command and Control server. Joanap is a 2-stage malware used to establish peer-to-peer communications and to manage botnets designed to enable other operations. It provides Hidden Cobra actors with the ability to exfiltrate data, drop and run secondary payloads, and initialize proxy communications on a compromised windows device where the communication is encrypted using *Rivest Cipher 4 encryption*. With the help of Joanap malware the Hidden Cobra actors could perform following activities on the compromised devices to the level of: (1) process management, (2) file and directory management, and (3) node management. Joanap typically infects a system as a file dropped by dropper malware (Destover, in this case) which users unknowingly downloaded either when they visit compromised sites or when they open malicious emails with attachments or web links [1].

## Dynamic DNS

Dynamic DNS allows for easier updating of DNS records

A key aspect of DDNS service is that changes to the IP assignments can be set to quickly propagate across the internet

A DDNS client resides either at a DDNS-enabled home router or the host server

DDNS service providers: DynDNS, No-IP, Securepoint DynDNS, Dynu

***Dynamic DNS*** or ***DDNS*** allows for *almost real-time, automatic updating of a name server in the DNS*, with active DDNS configured with hostnames, IP addresses and related information. Hence DDNS is a service offered in which the provider will allow users to control the IP address assignment of a domain. The user can access this IP address assignment through the provider and make changes as needed. One of the key aspects of a DDNS service (compared to a traditional DNS service) is that changes to the IP assignments can be set to quickly propagate across the internet, while a traditional DNS service may take longer to propagate or update various sources where a computer might seek to "look up" or resolve a domain [4].

In the networks using Dynamic DNS, when the client machine boots it gets an IP address from the DHCP and then updates the DNS records with the new IP obtained. This routine of updating the DNS records with the new IP address is done by a Dynamic DNS module residing either in the home router or the client machine. Such a feature allowing simple registration of dynamic IP address adds to the convenience of system/network intruders, to commit internet attacks and remain untraceable or go absconding.

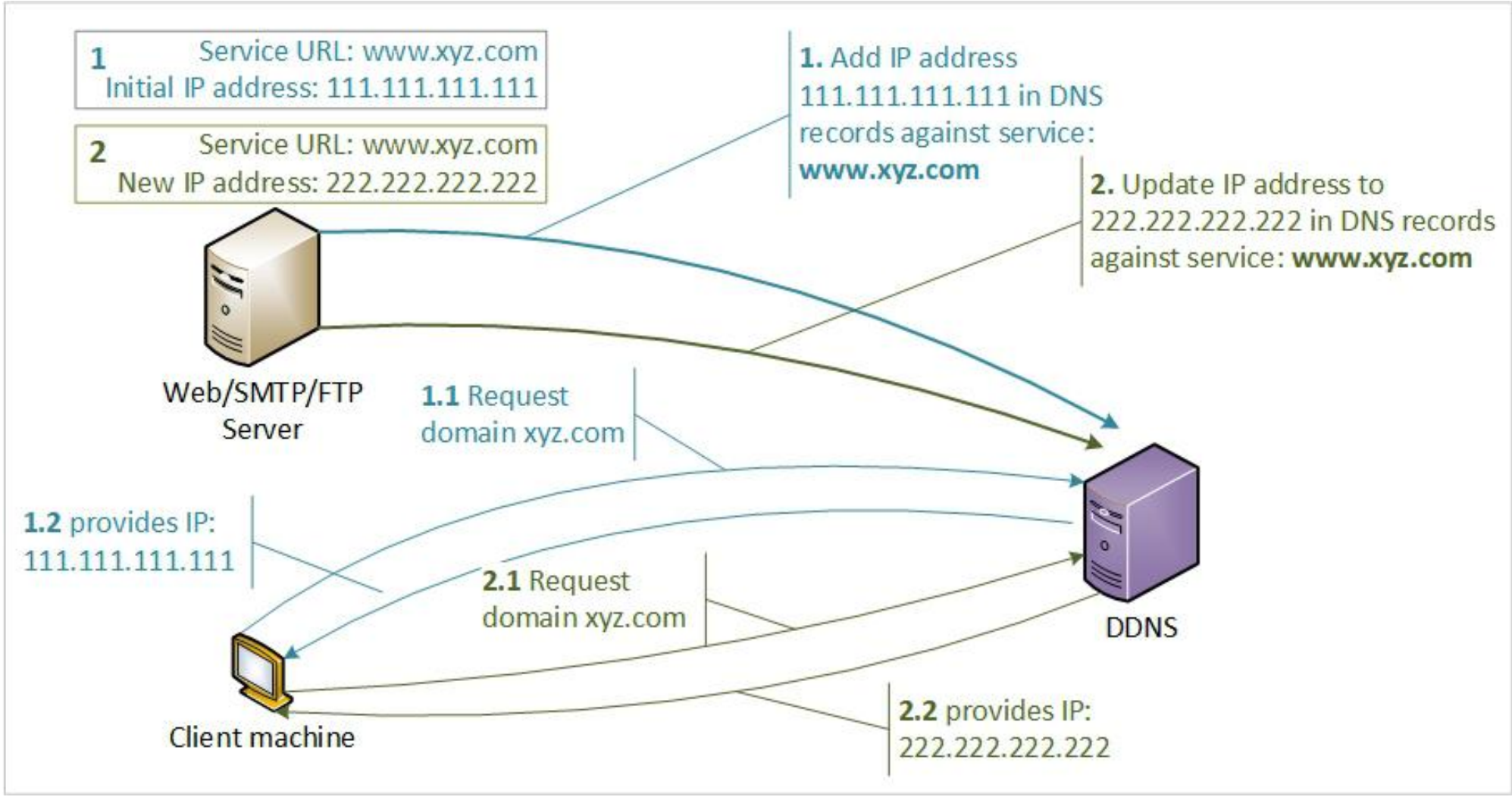

Figure 1: A representation of a dynamic update of DNS records in DDNS

To avail a DDNS service, one needs to sign up with a dynamic DNS service provider and install the DDNS client either in their DDNS-enabled home router or the host computer (the server, service provider). The DDNS client keeps monitoring the IP address and updates any change in the IP address to its DDNS service. This implies that as long as the DDNS client does its task, your domain name will continue to direct visitors to your host no matter how many times its IP address changes.

## Proxy Services

Proxy services are sometimes used as anonymizing service

A **proxy service** is provided by a ***proxy server***, which shows its IP address to the external networks while concealing the IP address of the machine from which the request originated. The proxy server acts as an intermediary for requests from clients seeking resources from servers in other networks.

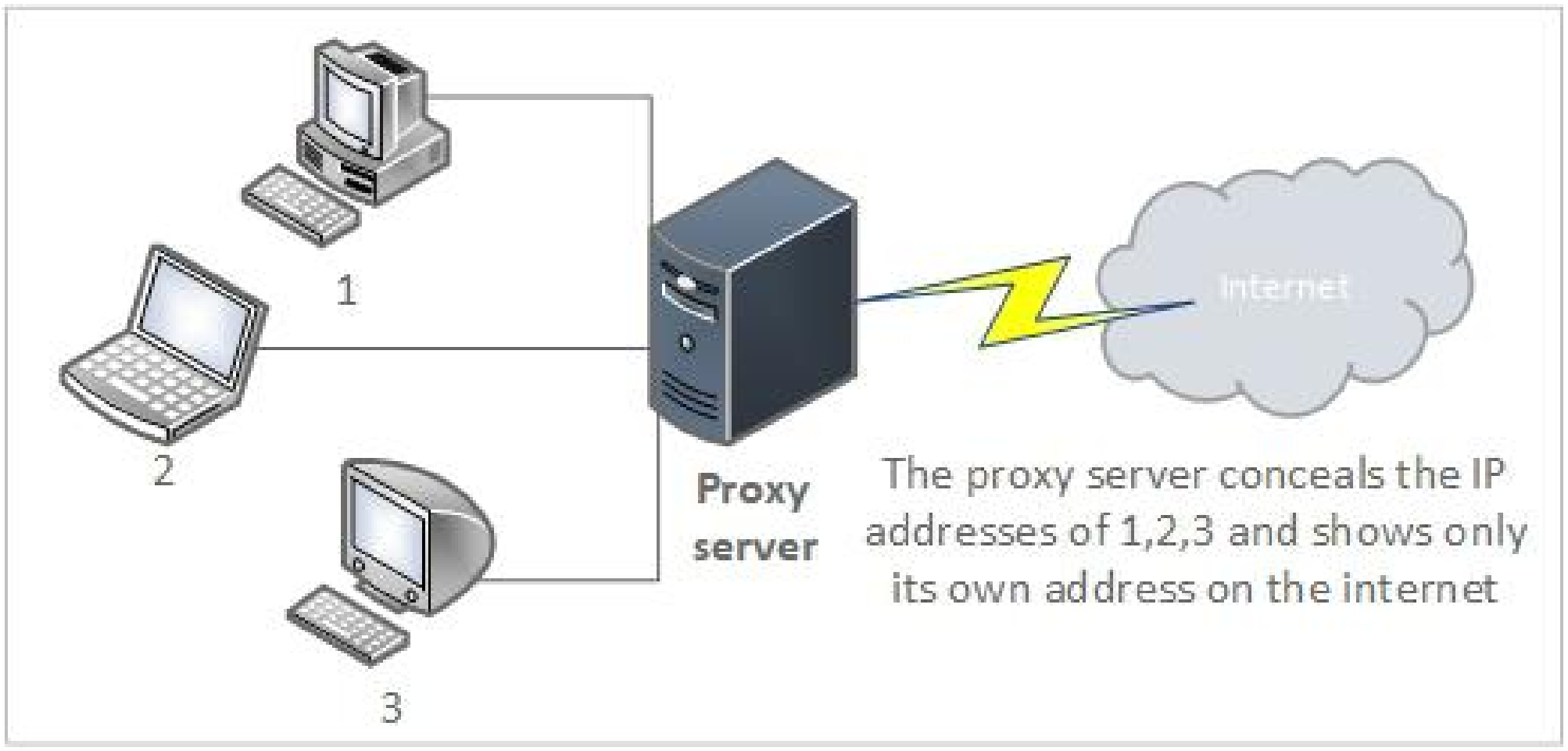

Figure 2: Working of a proxy server

Reconnaissance is the period of preliminary survey.

Such proxy services are often used by intruders as anonymizing services that can be used as a relay to conceal one's true IP address, and thus one's location. When such a service is used, the website being visited only "sees" the IP address of the proxy, not the user's true "home" IP address.

## The Intrusion

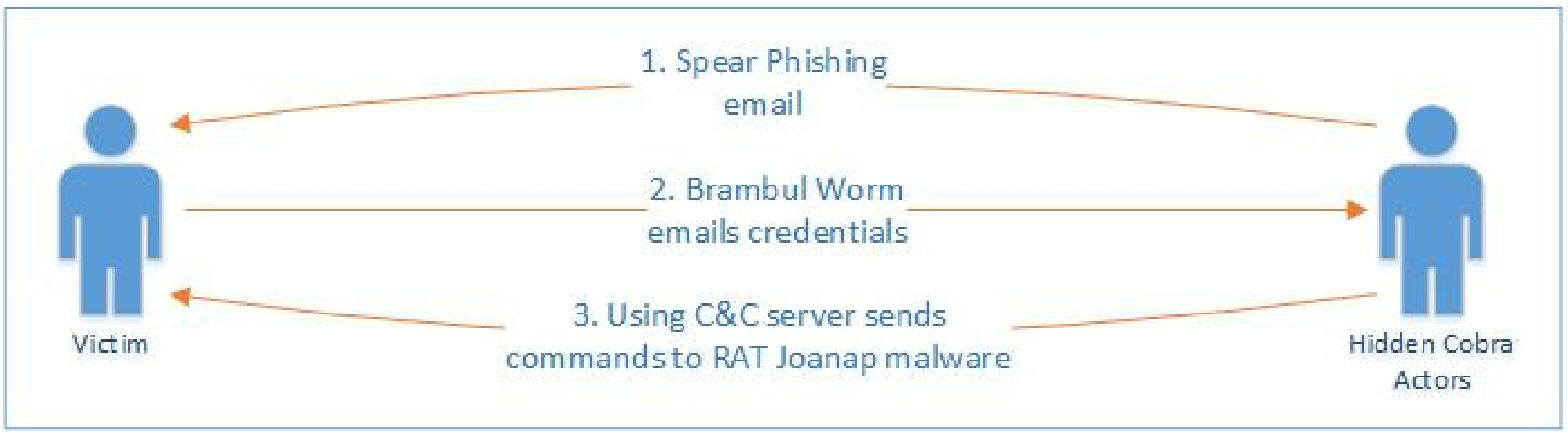

Figure 3: Logical representation of the intrusion

Phishing or Spear-phishing is an email-spoofing attack where the message appears to have come from a trusted source but actually its not

1) As in any intrusion, this one too was preceded by a period of reconnaissance, during which the Hidden Cobra actors surveyed the Internet and social media activities of their target victims. The results of the **reconnaissance** were used by the Hidden Cobra actors to perform **spear-phishing** attack using messages inviting the victims' interest. As an outcome of which the Hidden Cobra actors either stole the user credentials to install malware or directly installed malware on to the victims' machines.

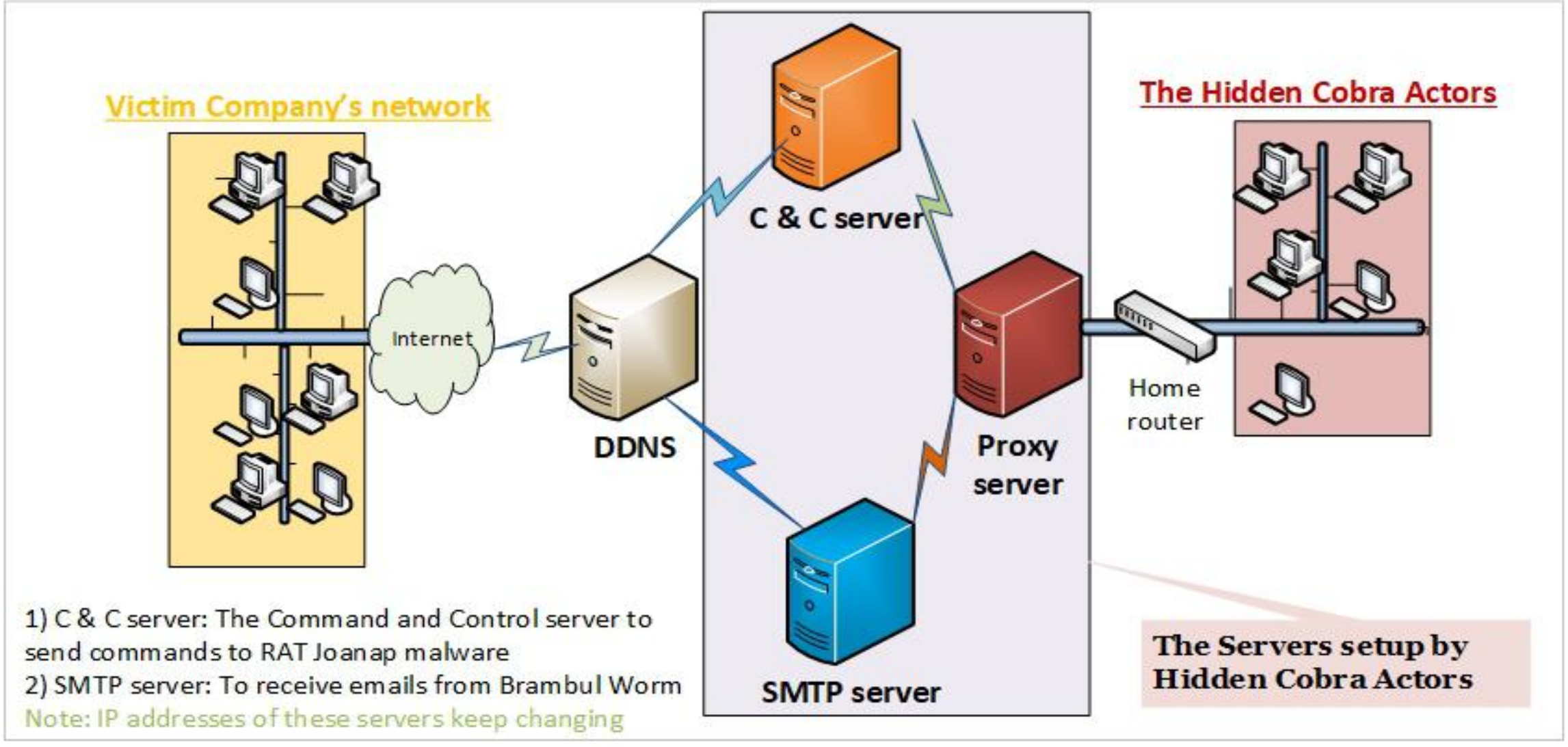

Figure 4: A representation of the possible network architecture of the whole intrusion event

*The Hidden Cobra Actors set-up an infrastructure comprising of SMTP server, C&C server and Proxy server.*

*Modus operandi included Spear Phishing, Brambul Worm, RAT Joanap malware, proxy services and the SMTP and C&C servers registered with DDNS*

2) The **Brambul worm** seated on the victims' machines then crawled through the network by exploiting the SMB protocol by brute force using the list of hard-coded passwords available with it, and gathered details of users and systems; which it sent via hard-coded email ids to Hidden Cobra actors through an SMTP server hosted by the actors themselves and registered with a DDNS service. By this, the Hidden Cobra actors were able to get access to more and more machines in the compromised networks. The SMTP could change its IP frequently as it was registered with DDNS, hence tracking the SMTP server and its location became challenging [3].

3) The Hidden Cobra Actors used a **Command and Control (C&C) server** and send commands to **RAT Joanap malware** installed in the victims' machines to control their activities. The C&C server hosted by them too was registered with a DDNS service.

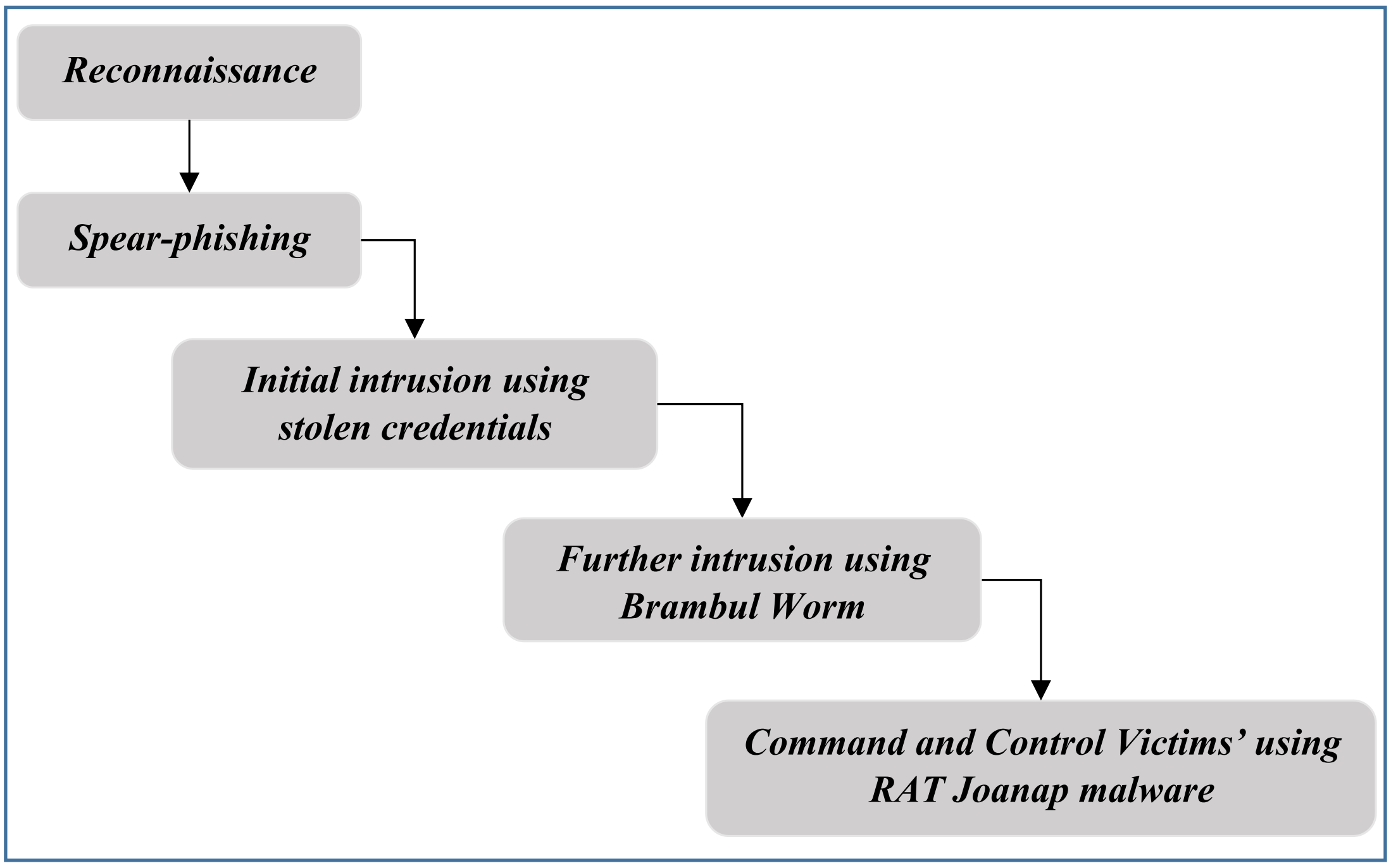

Figure 5: A flow chart representation of the intrusion events

## Malware Analysis Report

The Malware is a portable executable dll file implemented in VC++ for MS Windows

**Name**: 4731CBAEE7ACA37B596E38690160A749
**Size**: 208896 bytes
**Type**: PE32 executable (GUI) Intel 80386, for MS Windows
**MD5**: 4731cbaee7aca37b596e38690160a749
**SHA1**: 80fac6361184a3e24b33f6acb8688a6b7276b0f2
**SHA256**:
077d9e0e12357d27f7f0c336239e961a7049971446f7a3f10268d9439ef67885
**SHA512**:
9fdc1bf087d3e2fa80ff4ed749b11a2b3f863bed7a59850f6330fc1467c38eed05
2eee0337d2f82f9fe8e145f68199b966ae3c08f7ad1475b665beb8cd29f6d7
**Entropy**: 7.731026
**Description**:
This 32-bit Windows executable file drops two malicious applications. The first (a1c483b0ee740291b91b11e18dd05f0a460127acfc19d47b446d11cd0e26d717) is a fully functioning RAT. This malware has been identified as a RAT, providing a remote actor with the ability to exfiltrate data, drop and run secondary payloads, and provide proxy capabilities on a compromised Windows device. The malware binds to port 443 and listens for incoming connections from a remote operator, using the Rivest Cipher 4 (RC4) encryption algorithm to protect communications with its C&C. The second application (ea46ed5aed900cd9f01156a1cd446cbb3e10191f9f980e9f710ea1c20440c781) is a SMB worm that will spread to local subnets and external networks. The malware also creates a log entry in a file named "***mssscardprv.ax***", located in the %WINDIR%\system32 folder. The log entry includes the victim's Internet Protocol (IP) address, host name, and current system time [2].

## Mitigation plan

- Maintain up-to-date antivirus signatures and engines
- Keep operating system patches up-to-date
- Disable File and Printer sharing services. If these services are required, use strong passwords or Active Directory authentication
- Restrict users' ability (permissions) to install and run software
- Enforce a strong password policy and regular password changes
- Exercise caution when opening e-mail attachments
- Enable a personal firewall on agency workstations
- Disable unnecessary services on agency workstations and servers
- Scan for and remove suspicious e-mail attachments
- Monitor users' web browsing habits; restrict access to sites with unfavorable content
- Exercise caution when using removable media
- Scan all software downloaded from the Internet prior to executing
- Maintain situational awareness of the latest threats and implement appropriate ACLs [2]

## Acknowledgements

We express our sincere thanks to National Internet Exchange of India (NIXI) and Ministry of Electronics & Information Technology (MieTY) for their continued support.

## About the Authors

- **Gopinath Palaniappan,** Principal Technical Officer at CDAC, Bangalore, a researcher in Malware
- **Dr. Balaji Rajendran,** Joint Director at CDAC, Bangalore, a researcher in Cyber Security
- **Dr. S Sangeetha,** Assistant Professor at NIT,Tiruchirapalli, a researcher in NLP
- **Dr. Kumari Roshni V S**, Senior Director at CDAC, Bangalore, a researcher in Image processing